\documentclass[11pt]{article}
\usepackage[font=small]{caption}
\usepackage{tikz}
\usetikzlibrary{matrix,positioning}
\usepackage{authblk}

\usepackage{geometry}
\usepackage[english]{babel}
\usepackage[utf8]{inputenc}
\usepackage[T1]{fontenc}
\usepackage{indentfirst}
\usepackage{amsmath}
\usepackage{amssymb}
\usepackage{amsthm}
\usepackage{proof}
\usepackage{eufrak}
\usepackage{graphicx}
\usepackage{psfrag}

\allowhyphens

\newcommand{\UU}{\mathcal{U}}
\newcommand{\sech}{\text{sech}}

\newcommand{\egesz}{\mathbb{Z}}

\newcommand{\ordo}{\mathcal{O}}
\newcommand{\Pe}{\mathcal{P}}
\newcommand{\La}{\mathcal{L}}

\newcommand{\ket}[1]{{\left|#1\right\rangle}}

\definecolor{mycolor}{rgb}{0.9,0.9,0.8}

\tikzset{
   mycell/.style={draw, minimum size=0.3cm},
 }

 \tikzset{
     dot2/.style={mycell,
       append after command={
   \pgfextra \fill[mycolor] (\tikzlastnode) (-0.15cm,-0.15cm) rectangle (0.15cm,0.15cm); \endpgfextra
         \pgfextra \fill (\tikzlastnode) circle[radius=0.08cm]; \endpgfextra}}
   }

 \tikzset{
     dot/.style={mycell,
       append after command={
           \pgfextra \fill (\tikzlastnode) circle[radius=0.08cm]; \endpgfextra}}
     }   

 \tikzset{
     rec/.style={mycell,
       append after command={
   \pgfextra \fill[mycolor] (\tikzlastnode) (-0.15cm,-0.15cm) rectangle (0.15cm,0.15cm); \endpgfextra}
      }
   }
     

\usepackage{ifpdf}

\ifpdf
\usepackage{epstopdf}
\usepackage[pdftex,colorlinks,urlcolor=blue,citecolor=blue,linkcolor=blue]{hyperref}
\else
\usepackage[hypertex,colorlinks,urlcolor=blue,citecolor=blue,linkcolor=blue]{hyperref}
\fi
\begin{document}
\numberwithin{equation}{section}

\title{A Yang-Baxter integrable cellular automaton \\ with a four site update rule}
  \author[1]{Bal\'azs Pozsgay}
  \affil[1]{MTA-ELTE “Momentum” Integrable Quantum Dynamics Research Group, \protect\\ Department of Theoretical Physics, \protect\\ Eötvös
  Loránd University}

  \maketitle

  \abstract{We present a one dimensional reversible block cellular automaton, where the time evolution is dictated by a period 3
    cycle of update rules. At each time step a subset of the cells is updated using a four site rule with two control bits and
    two action bits. The model displays rich dynamics. There are three types of stable particles, left
    movers, right movers and ``frozen'' bound states that only move as an effect of scattering with the left and right
    movers.
    Multi-particle scattering in the system is factorized.  We  embed the model into the canonical framework of
    Yang-Baxter integrability  
by    rigorously proving the existence of a commuting set of diagonal-to-diagonal transfer matrices. The construction
    involves a new type of Lax operator.
}

\section{Introduction}

Cellular automata are discrete models of classical computation, that can describe complex systems in various domains.
They can be applied in physics and chemistry (modeling for example fluid flow, earthquakes, or galaxy formation), but
also in biology for the modeling of pattern formation \cite{cellaut-review,wolfram-book}. Their beauty lies in the
simplicity of their rules, which 
can nevertheless lead to very complex behaviour. A famous example is Conway's Game of Life, which is known to be Turing complete
\cite{rendell-game-of-life-book}. 

In physics the cellular automata can play the role of ``toy models of nature'': they have the potential to make
a bridge between the microscopic and macroscopic physical laws. An important example is the desired rigorous derivation of
statistical physics and thermodynamics from the microscopic rules. A special class of models that have been
studied for this purpose are the one dimensional reversible cellular automata \cite{wolfram-cellaut1,cellaut-class}. The
most often studied models are the elementary automata, where the cells are organized in a row, each cell has two states
(occupied or empty), the update 
rules are translationally invariant, and the next state of each cell depends on the current state of the cell and its
immediate neighbours. Counting all the possibilities it can be seen that there are $2^8=256$ different update rules in
this class, and they have been studied and classified in \cite{wolfram-cellaut1,cellaut-class}. Reversible models
defined on a light cone lattice were afterwards classified and studied in \cite{rule54}.
If cellular automata can serve as toy models for nature, it is very natural to ask whether they can lead to so-called integrable
models with non-trivial dynamics. The answer is yes, as we discuss below.

One dimensional integrable models are special systems where the dynamical
processes are constrained, leading to completely elastic and factorized scattering
\cite{sutherland-book,Baxter-Book,mussardo-review}. In the quantum mechanical case the wave functions can be constructed
with exact methods, leading eventually to the study of the equilibrium and non-equilibrium dynamics of these
systems. One of the long standing problems is to compute the macroscopic transport properties from the microscopic
dynamics, 
and this area witnessed tremendous progress since the development of Generalized Hydrodynamics (GHD)
\cite{doyon-ghd,jacopo-ghd,ghd-misc}.  Yet, despite this progress relatively few elements of the theory are
rigorously proven (for various partial results see for example
\cite{doyon-hydro-proj,granet-essler-ghd,sajat-currents-review,katja-bruno-rule54-ghd}). There
remains a search for toy models, which are simple enough so that the exact real time dynamics can be computed,
but complicated enough in order to have non-trivial interaction effects.

One such toy model is the so-called Rule54 model, which is sometimes also called the Floquet Fredrickson-Andersen (FFA)
model, and often claimed to be the simplest interacting integrable model.  It is a reversible cellular automaton proposed in \cite{rule54}, where it was shown to have stable particles with soliton-like behaviour. The
particles can be right movers or left movers which propagate with a velocity of $\pm 1$. The left-right scattering
events are non-trivial: they lead to a displacement of the particle orbits. The Rule54 model has been the subject of
active research in the last couple of years, see the recent review \cite{rule54-review}.
Exact real time evolution in the Rule54 model was computed for special initial states in the recent works
\cite{katja-bruno-lorenzo-rule54,katja-bruno-rule54-ghd,rule54-entangl}, leading to a rigorous proof of the predictions of GHD in
special cases.

Up to now the Rule54 model remained a somewhat isolated example, residing on the intersection of the worlds of cellular
automata and 
integrable models. Furthermore, the connections with the standard methods of integrability remained obscure. 
A quantum mechanical deformation of the Rule 54 model was presented in \cite{vasseur-rule54}, where the so-called Bethe
Ansatz solution of the model was also given. Nevertheless it was not known how to embed the model into the famous
Yang-Baxter algebra, which is known to be central to integrable models. It is remarkable that the exact computations  of
\cite{katja-bruno-lorenzo-rule54,katja-bruno-rule54-ghd} did not use any of 
the standard methods of integrability, and instead they were built on methods developed for ``dual unitary quantum
gate'' models, see \cite{dual-unitary-0,prosen-dual1,dual-unitary-2}. The 
dual unitary models are not integrable in the traditional sense, but they allow for exact solutions
\cite{prosen-dual1,dual-unitary-2}. 

In this paper it is our goal to construct new cellular automata that are integrable, that show solitonic
dynamics, and we  make a connection with the canonical methods of quantum integrability.

The possibilities for 2-state cellular automata with strictly local update rules which are also strictly homogeneous in
space and time are exhausted by the 
classifications of \cite{wolfram-cellaut1,cellaut-class,rule54}. Thus new models can be found if we relax at least some of the
requirements. Staying in the realm of 2-state systems one possibility is to look for block cellular
automata, where the cells are updated in blocks, and the 
grouping of the cells into blocks changes periodically with each time step. Such a cellular automaton is a
classical version of the brickwork-type local quantum circuits used often in quantum computation.
Integrable quantum gate models were studied for example in \cite{integrable-trotterization}, where a period 2 Floquet
cycle of two-site quantum gates was used as an integrable Trotterization of the XXZ Heisenberg chains. It is very
natural to attempt to 
construct block cellular automata using this setup. However, the only deterministic case of the models of
\cite{integrable-trotterization} is trivial: it leads to simple permutations of quantum spaces and thus to free particle
propagation for the resulting cellular automaton. We conclude that new models (with spin-1/2 variables) can be found
only if we consider block sizes bigger than 2.

Therefore we consider block cellular automata with block sizes of 3
and 4. This gives us enough freedom to accommodate new and non-trivial solitonic dynamics. The setup we find is similar to that
of \cite{integrable-trotterization}, but now we apply a period 3 Floquet cycle of block updates. We obtain a new
integrable cellular automaton, which is perhaps the ``next simplest'' integrable model after the Rule54 model. The
origins of this model lie in the recent works \cite{fracton1,folded1,folded2,sajat-folded}, which treated the so-called
folded XXZ model, a spin-1/2 chain with remarkably simple dynamical properties. In turn, the folded XXZ model is a
special case of the Bariev model \cite{bariev-model}, which can be interpreted also as a zig-zag spin ladder.
As we explain below, our cellular
automaton can be regarded as a classical version of the folded XXZ model, and thus of the Bariev model.

In Section \ref{sec:4} we introduce our main model, which uses a 4-site update rule. An alternative version with a
3 site rule (also called the bond model) is introduced in Section \ref{sec:3}. We present a quantum mechanical extension
(a brickwork-type quantum circuit) in Section \ref{sec:gates}. The integrability of the models is established in Section
\ref{sec:Lax}, where a set of local conserved charges is derived for a certain diagonal-to-diagonal transfer
matrix. Our conclusions and a list of open problems is given in Section \ref{sec:conclude}.

\section{The cellular automaton: a 4 site model}

\label{sec:4}

Consider a spin-1/2 chain of length $L$ with periodic boundary conditions. We require that $L=3k$ with $k\in\egesz$.
Let us denote the basis states of a local space as $\ket{\circ}$ and $\ket{\bullet}$; they correspond to the up and down
spins in the usual $S^z$  basis. Correspondingly we introduce the local projection operators onto these basis states:
\begin{equation}
  P^\circ=\frac{1+\sigma^z}{2},\qquad   P^\bullet=\frac{1-\sigma^z}{2}.
\end{equation}
Here $\sigma^z$ is the standard Pauli matrix. We will also use the standard raising and lowering operators with a single
matrix element given by the action
\begin{equation}
  \sigma^- \ket{\circ}=\ket{\bullet},\qquad \sigma^+\ket{\bullet}=\ket{\circ}.
\end{equation}

Below we introduce quantum circuits, which update the state of the model in discrete time steps. In our main example the
update is deterministic. This means that the classical configurations (states of the computational basis) remain
classical after each step, and linear combinations do not arise. In this case the quantum circuit model can be
interpreted as a cellular automaton.

We start with the unitary quantum gate $U^{(4)}(j)=U^{(4)}_{j,j+1,j+2,j+3}$ which acts on the 4 sites of the segment
$[j,\dots,j+3]$. Its explicit form is 
\begin{equation}
  \label{U4}
  U^{(4)}(j)=\left(P^\bullet_j P^\bullet_{j+3}+P^\circ_j P^\circ_{j+3}  \right) \Pe_{j+1,j+2}
  +P^\bullet_j P^\circ_{j+3}+P^\circ_j P^\bullet_{j+3},
\end{equation}
where $\Pe_{j+1,j+2}$ is the permutation operator acting on two sites.
This is a unitary operator, which is in fact deterministic in the computational basis: acting on any basis state of a
segment of 4 sites it results in a single basis state with coefficient 1.

The interpretation of this unitary gate is the
following. The outer two spins on site $j$ and $j+3$ are control bits, which control the operation on the inner two bits. If the
state of the control bits is the same (be it $\circ$ or $\bullet$) then $U^{(4)}(j)$ permutes the Hilbert spaces on sites
$j+1$ and $j+2$. If the two control bits have different values then $U^{(4)}(j)$ acts as identity on the action bits.

The non-zero off-diagonal elements $U^{(4)}(j)$ correspond to the moves shown below:
\begin{center}
  \begin{tikzpicture}
 
\matrix (m) [matrix of nodes, row sep=-\pgflinewidth, column sep=-\pgflinewidth, 
    nodes={mycell}, nodes in empty cells]
    {
     && |[dot]| & \\
};
\node at (1,0) {$\leftrightarrow$};
\node at (7,0) {$\leftrightarrow$};
\node at (4,0) {and};
\begin{scope}[xshift=2cm]
\matrix (m) [matrix of nodes, row sep=-\pgflinewidth, column sep=-\pgflinewidth, 
    nodes={mycell}, nodes in empty cells]
   {
    & |[dot]| &&  \\
};
\end{scope}

\begin{scope}[xshift=6cm]
\matrix (m) [matrix of nodes, row sep=-\pgflinewidth, column sep=-\pgflinewidth, 
    nodes={mycell}, nodes in empty cells]
   {
 |[dot]| & |[dot]| && |[dot]| \\
};
\end{scope}
\begin{scope}[xshift=8cm]
\matrix (m) [matrix of nodes, row sep=-\pgflinewidth, column sep=-\pgflinewidth, 
    nodes={mycell}, nodes in empty cells]
   {
 |[dot]| & |[dot]| && |[dot]| \\
};
\end{scope}
  \end{tikzpicture}
\end{center}
The action of $U^{(4)}(j)$ leaves all other 4 site combinations invariant.

It is important that $U^{(4)}(j)$ is spin-flip invariant. Furthermore, it is involutive:
\begin{equation}
  \left(U^{(4)}(j)\right)^2=1.
\end{equation}

Now we construct the update rules for the cellular automaton. The idea is to construct a period 3 Floquet update rule,
such that 3 different update steps are applied periodically. In each update step we identify a subset of the bits as
control bits: we designate every third bit as a control bit and apply the 4-site update rule on the
pairs of action bits between each pair of neighbouring control bits. In the next time step we shift the selection of the control
bits to the right by one site.

This rule can be formalized as follows. We introduce the single step operators $V_m$ with $m=1,2,3$ as
\begin{equation}
  \label{Vm}
  V_m=\prod_{j=1}^{L/3} U^{(4)}(3j+m).
\end{equation}
Here the $U^{(4)}$-operators are positioned 3 sites apart, which means that two neighbouring $U^{(4)}$ operators share a 
control bit. The control bits are not changed during the update, thus the operators in the above product commute with
each other. The $V_m$ operators with different values of $m$ are related by simple translation on the lattice: they merely
correspond to shifted choices for the control bits.

Finally we define the period 3 Floquet operator $\mathcal{V}$ as
\begin{equation}
  \label{V}
  \mathcal{V}=V_3V_2V_1.
\end{equation}
The time evolution in the cellular automaton is then generated by a repeated action of $\mathcal{V}$ on the chain.

It is important that the operators $V_m$ with different values of $m$ do not commute with each other, and it is
important to have a pre-defined order of their action. A different ordering leads to a different model.

The single step update rules are reversible, thus the cellular automaton is reversible. Due to the involutive property
of the 4-site gate the inverse of the Floquet operator is given by 
\begin{equation}
(  \mathcal{V})^{-1}=\mathcal{V}'=V_1V_2V_3.
\end{equation}
This ordering of the single step operators corresponds to shifting the control bits to the left after each update
step. Thus $\mathcal{V}'$ is 
also the space reflected version of our model, which also implies that the model is $PT$ symmetric. 

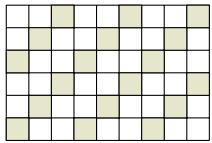
\begin{figure}[b!]
  \centering
\begin{tikzpicture}
\draw[step=0.3cm] (0,0) grid (2.7,1.8);

\foreach \x in {1,...,3}
{
  \foreach \z in {0,1}
  {
    \foreach \y in {1,...,3}
    {
      \begin{scope}[xshift=(3*(\x-1)+\y-1)*0.3cm,yshift=(\y-1+\z*3)*0.3cm]
        \filldraw[fill=mycolor] (0,0) rectangle (0.3,0.3);      
      \end{scope}
     }
  }  
  }
\end{tikzpicture}
\caption{An example for the placement of the control bits on a rectangular lattice with $L=9$ and two complete Floquet
  cycles of length $3$. The vertical axis corresponds to time: each row will depict the state of the cells at a given
  time. The filled cells are the control bits, that influence the update for the action bits between
  them. Periodic boundary conditions are understood.}
\label{fig:coloring}
\end{figure}

Below we show concrete examples for the dynamics of the model. The time evolution is presented on a two dimensional
lattice, where time flows in the vertical direction, such that the bottom line is the initial condition. Sites are
indexed such that the leftmost cell corresponds to  $j=1$, and the site coordinates increase to the right. We always
assume periodic boundary conditions in the horizontal direction, but this is not used explicitly in the
examples. The control bits for the update steps are placed according to the formulas \eqref{V}-\eqref{Vm}. This means
that in the figures the bottom left cell is always a control bit for the corresponding update step. The other control
bits are placed on parallel lines in a diagonal direction. We demonstrate this in  Figure \ref{fig:coloring}, which
shows an empty lattice (all sites unoccupied) with a background coloring of the  control bits. We keep this coloring
also in the other Figures to be 
presented below, such that the time evolution can be easily followed/checked by the reader.

\subsection{Dynamics}

Now we investigate the dynamics of this cellular automaton. We consider various scenarios (initial states) and look at
the outcome of the time evolution. We simply just run the code of the cellular automaton on a periodic lattice and we
display the outcome of the dynamics on a two dimensional graph, where the vertical axis corresponds to the time
variable. These small runs can be considered as ``experiments'' with the cellular automaton.

First we consider single particles: we take initial
conditions where a single particle is embedded into a vacuum of empty states. Note that due to spin flip invariance
the embedding of a single empty state into a fully occupied sea of particles would lead to the same dynamics.

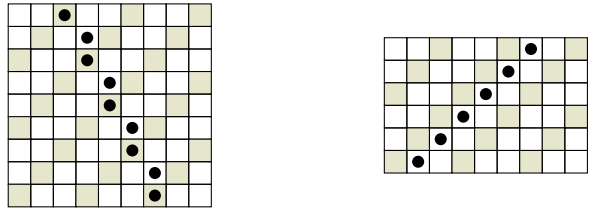
\begin{figure}[b!]
  \centering
  \begin{tikzpicture}

\matrix (m) [matrix of nodes, row sep=-\pgflinewidth, column sep=-\pgflinewidth, 
    nodes={mycell}, nodes in empty cells]
    {
      && |[dot2]| & &&|[rec]|  &&&|[rec]|\\ 
&|[rec]|  && |[dot]| & |[rec]|  &&&|[rec]|  &\\ 
|[rec]|  &&& |[dot2]| & &&|[rec]|  &&\\ 
&&|[rec]|  && |[dot]| & |[rec]|  &&&|[rec]|\\ 
&|[rec]|  &&& |[dot2]| & &&|[rec]|  &\\ 
|[rec]|  &&&|[rec]|  && |[dot]| & |[rec]|  &&\\ 
&&|[rec]|  &&& |[dot2]| & &&|[rec]|\\ 
&|[rec]|  &&&|[rec]|  && |[dot]| & |[rec]|  &\\ 
|[rec]|  &&&|[rec]|  &&& |[dot2]| & &\\ 
};

\begin{scope}[xshift=5cm]
  \matrix (m) [matrix of nodes, row sep=-\pgflinewidth, column sep=-\pgflinewidth, 
    nodes={mycell}, nodes in empty cells]
    {
&&|[rec]|  &&&|[rec]|  & |[dot]| & &|[rec]|\\ 
&|[rec]|  &&&|[rec]|  & |[dot]| & &|[rec]|  &\\ 
|[rec]|  &&&|[rec]|  & |[dot]| & &|[rec]|  &&\\ 
&&|[rec]|  & |[dot]| & &|[rec]|  &&&|[rec]|\\ 
&|[rec]|  & |[dot]| & &|[rec]|  &&&|[rec]|  &\\ 
|[rec]|  & |[dot]| & &|[rec]|  &&&|[rec]|  &&\\ 
};
\end{scope}

\end{tikzpicture}
  \caption{Propagation of a left mover and a right mover.}
  \label{fig:1pt}
\end{figure}

The results of single particle propagation are depicted in
Fig. \ref{fig:1pt}.
We find that there are two types of
particles in the system: right movers and left movers. The position of the particle with respect to the control
bits in the given time step decides whether it is a right or a left mover. The rules are the following. Let us assume
that at a given time the particle is within the segment $j\dots j+3$ as the unitary gate $U^{(4)}(j)$ is applied.
If the particle is at position $j+1$ ($j+2$) then it is moved to the right (left), respectively. If the particle
is at site $j$ or $j+4$ then it is not moved in the current step, but it becomes a left mover in the next time step,
because it is moved in the next step either by $U^{(4)}(j-2)$ or by $U^{(4)}(j+1)$. Continuing the time evolution we observe that
a right mover 
stays a right mover and it has a constant velocity of 1. Similarly, a left mover stays a left mover, but it moves only
in every second step, thus it has an average velocity of $-1/2$.

Looking at the update rules we also observe that blocks of particles with a minimum length of 2 are
frozen, they do not propagate. An example with length 2 is shown below. 
\begin{center}\begin{tikzpicture}
  
\matrix (m) [matrix of nodes, row sep=-\pgflinewidth, column sep=-\pgflinewidth, 
    nodes={mycell}, nodes in empty cells]
    {
& & &  |[dot]| &  |[dot]| & & & \\
    };
  \end{tikzpicture}
\end{center}
These configurations are interpreted as bound states that are not dynamical. The absence of their propagation
is a result of the control bits: none of the $U^{(4)}(j)$ gates moves any of the particles in such configurations.

We have thus essentially 3 types of particles in the system with 3 different speeds: right movers with speed 1, left
movers with speed -1/2, and bound states with speed 0. It can be seen that bound states with different length behave in the
same way, thus there is no need to treat them as separate particle species.

Let us now consider the scattering of particles. 

\begin{figure}[b!]
  \centering
  \begin{center}\begin{tikzpicture}

      
\matrix (m) [matrix of nodes, row sep=-\pgflinewidth, column sep=-\pgflinewidth, 
    nodes={mycell}, nodes in empty cells]
    {
&& |[dot2]| & &&|[rec]|  &&&|[rec]|  &&&|[rec]|  & |[dot]| & &|[rec]|\\ 
&|[rec]|  && |[dot]| & |[rec]|  &&&|[rec]|  &&&|[rec]|  & |[dot]| & &|[rec]|  &\\ 
|[rec]|  &&& |[dot2]| & &&|[rec]|  &&&|[rec]|  & |[dot]| & &|[rec]|  &&\\ 
&&|[rec]|  && |[dot]| & |[rec]|  &&&|[rec]|  & |[dot]| & &|[rec]|  &&&|[rec]|\\ 
&|[rec]|  &&& |[dot2]| & &&|[rec]|  & |[dot]| & &|[rec]|  &&&|[rec]|  &\\ 
|[rec]|  &&&|[rec]|  && |[dot]| & |[rec]|  & |[dot]| & &|[rec]|  &&&|[rec]|  &&\\ 
&&|[rec]|  &&& |[dot2]| & & |[dot]| & |[rec]|  &&&|[rec]|  &&&|[rec]|\\ 
&|[rec]|  &&&|[rec]|  & |[dot]| & & |[dot2]| & &&|[rec]|  &&&|[rec]|  &\\ 
|[rec]|  &&&|[rec]|  & |[dot]| & &|[rec]|  && |[dot]| & |[rec]|  &&&|[rec]|  &&\\ 
&&|[rec]|  & |[dot]| & &|[rec]|  &&& |[dot2]| & &&|[rec]|  &&&|[rec]|\\ 
&|[rec]|  & |[dot]| & &|[rec]|  &&&|[rec]|  && |[dot]| & |[rec]|  &&&|[rec]|  &\\ 
|[rec]|  & |[dot]| & &|[rec]|  &&&|[rec]|  &&& |[dot2]| & &&|[rec]|  &&\\ 
    };
  \end{tikzpicture}
\end{center}
  \caption[sc]{Scattering of a right mover and a left mover.}
  \label{fig:scatt1}
\end{figure}
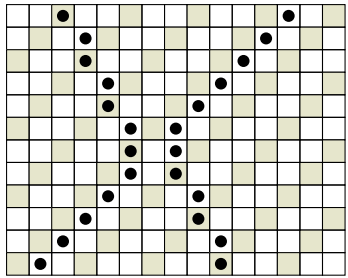

First we look at the scattering of a right mover and a left mover. It
is clear from the rules that as the two particles approach each other, there will not be a single time when the two
particles occupy neighbouring positions. Any hopping in- and out- of such configuration is forbidden by the control
bits. However, there is a non-trivial scattering between the two particles, which is depicted on Fig \ref{fig:scatt1}. We
see that the right mover eventually continues its orbit, but the left mover suffers a displacement: its outgoing orbit
is shifted by one lattice unit to the left and one lattice unit down. This is equivalent to a time delay of $3$ units,
which could be considered as a displacement of $\Delta x=-1.5$ due to the average speed of -1/2.

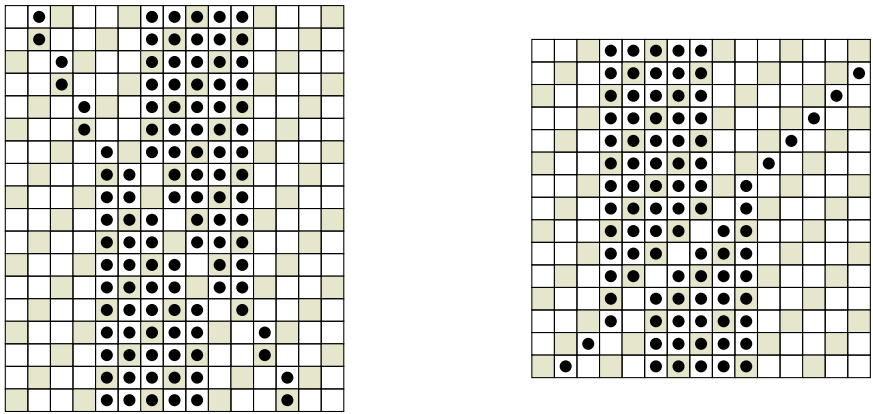
\begin{figure}[b!]
  \centering
  \begin{center}\begin{tikzpicture}
\matrix (m) [matrix of nodes, row sep=-\pgflinewidth, column sep=-\pgflinewidth, 
    nodes={mycell}, nodes in empty cells]
    {
& |[dot]| & |[rec]|  &&&|[rec]|  & |[dot]| &  |[dot]| &  |[dot2]| &  |[dot]| &  |[dot]| & |[rec]|  &&&|[rec]|\\ 
& |[dot2]| & &&|[rec]|  && |[dot]| &  |[dot2]| &  |[dot]| &  |[dot]| &  |[dot2]| & &&|[rec]|  &\\ 
|[rec]|  && |[dot]| & |[rec]|  &&& |[dot2]| &  |[dot]| &  |[dot]| &  |[dot2]| &  |[dot]| & &|[rec]|  &&\\ 
&& |[dot2]| & &&|[rec]|  & |[dot]| &  |[dot]| &  |[dot2]| &  |[dot]| &  |[dot]| & |[rec]|  &&&|[rec]|\\ 
&|[rec]|  && |[dot]| & |[rec]|  && |[dot]| &  |[dot2]| &  |[dot]| &  |[dot]| &  |[dot2]| & &&|[rec]|  &\\ 
|[rec]|  &&& |[dot2]| & && |[dot2]| &  |[dot]| &  |[dot]| &  |[dot2]| &  |[dot]| & &|[rec]|  &&\\ 
&&|[rec]|  && |[dot]| & |[rec]|  & |[dot]| &  |[dot]| &  |[dot2]| &  |[dot]| &  |[dot]| & |[rec]|  &&&|[rec]|\\ 
&|[rec]|  &&& |[dot2]| &  |[dot]| & & |[dot2]| &  |[dot]| &  |[dot]| &  |[dot2]| & &&|[rec]|  &\\ 
|[rec]|  &&&|[rec]|  & |[dot]| &  |[dot]| & |[rec]|  & |[dot]| &  |[dot]| &  |[dot2]| &  |[dot]| & &|[rec]|  &&\\ 
&&|[rec]|  && |[dot]| &  |[dot2]| &  |[dot]| & & |[dot2]| &  |[dot]| &  |[dot]| & |[rec]|  &&&|[rec]|\\ 
&|[rec]|  &&& |[dot2]| &  |[dot]| &  |[dot]| & |[rec]|  & |[dot]| &  |[dot]| &  |[dot2]| & &&|[rec]|  &\\ 
|[rec]|  &&&|[rec]|  & |[dot]| &  |[dot]| &  |[dot2]| &  |[dot]| & & |[dot2]| &  |[dot]| & &|[rec]|  &&\\ 
&&|[rec]|  && |[dot]| &  |[dot2]| &  |[dot]| &  |[dot]| & |[rec]|  & |[dot]| &  |[dot]| & |[rec]|  &&&|[rec]|\\ 
&|[rec]|  &&& |[dot2]| &  |[dot]| &  |[dot]| &  |[dot2]| &  |[dot]| & & |[dot2]| & &&|[rec]|  &\\ 
|[rec]|  &&&|[rec]|  & |[dot]| &  |[dot]| &  |[dot2]| &  |[dot]| &  |[dot]| & |[rec]|  && |[dot]| & |[rec]|  &&\\ 
&&|[rec]|  && |[dot]| &  |[dot2]| &  |[dot]| &  |[dot]| &  |[dot2]| & && |[dot2]| & &&|[rec]|\\ 
&|[rec]|  &&& |[dot2]| &  |[dot]| &  |[dot]| &  |[dot2]| &  |[dot]| & &|[rec]|  && |[dot]| & |[rec]|  &\\ 
|[rec]|  &&&|[rec]|  & |[dot]| &  |[dot]| &  |[dot2]| &  |[dot]| &  |[dot]| & |[rec]|  &&& |[dot2]| & &\\
};

\begin{scope}[xshift=7cm]
\matrix (m) [matrix of nodes, row sep=-\pgflinewidth, column sep=-\pgflinewidth, 
    nodes={mycell}, nodes in empty cells]
    {
&&|[rec]|  & |[dot]| &  |[dot]| &  |[dot2]| &  |[dot]| &  |[dot]| & |[rec]|  &&&|[rec]|  &&&|[rec]|\\ 
&|[rec]|  && |[dot]| &  |[dot2]| &  |[dot]| &  |[dot]| &  |[dot2]| & &&|[rec]|  &&&|[rec]|  & |[dot]| \\ 
|[rec]|  &&& |[dot2]| &  |[dot]| &  |[dot]| &  |[dot2]| &  |[dot]| & &|[rec]|  &&&|[rec]|  & |[dot]| & \\ 
&&|[rec]|  & |[dot]| &  |[dot]| &  |[dot2]| &  |[dot]| &  |[dot]| & |[rec]|  &&&|[rec]|  & |[dot]| & &|[rec]|\\ 
&|[rec]|  && |[dot]| &  |[dot2]| &  |[dot]| &  |[dot]| &  |[dot2]| & &&|[rec]|  & |[dot]| & &|[rec]|  &\\ 
|[rec]|  &&& |[dot2]| &  |[dot]| &  |[dot]| &  |[dot2]| &  |[dot]| & &|[rec]|  & |[dot]| & &|[rec]|  &&\\ 
&&|[rec]|  & |[dot]| &  |[dot]| &  |[dot2]| &  |[dot]| &  |[dot]| & |[rec]|  & |[dot]| & &|[rec]|  &&&|[rec]|\\ 
&|[rec]|  && |[dot]| &  |[dot2]| &  |[dot]| &  |[dot]| &  |[dot2]| & & |[dot]| & |[rec]|  &&&|[rec]|  &\\ 
|[rec]|  &&& |[dot2]| &  |[dot]| &  |[dot]| &  |[dot2]| & & |[dot]| &  |[dot2]| & &&|[rec]|  &&\\ 
&&|[rec]|  & |[dot]| &  |[dot]| &  |[dot2]| & & |[dot]| &  |[dot2]| &  |[dot]| & &|[rec]|  &&&|[rec]|\\ 
&|[rec]|  && |[dot]| &  |[dot2]| & & |[dot]| &  |[dot2]| &  |[dot]| &  |[dot]| & |[rec]|  &&&|[rec]|  &\\ 
|[rec]|  &&& |[dot2]| & & |[dot]| &  |[dot2]| &  |[dot]| &  |[dot]| &  |[dot2]| & &&|[rec]|  &&\\ 
&&|[rec]|  & |[dot]| & & |[dot2]| &  |[dot]| &  |[dot]| &  |[dot2]| &  |[dot]| & &|[rec]|  &&&|[rec]|\\ 
&|[rec]|  & |[dot]| & &|[rec]|  & |[dot]| &  |[dot]| &  |[dot2]| &  |[dot]| &  |[dot]| & |[rec]|  &&&|[rec]|  &\\ 
|[rec]|  & |[dot]| & &|[rec]|  && |[dot]| &  |[dot2]| &  |[dot]| &  |[dot]| &  |[dot2]| & &&|[rec]|  &&\\ 
    };  
\end{scope}

  \end{tikzpicture}
\end{center}
\caption[sc]{Scattering of a single particle and a bound state of length 5. Figure on the left: incoming left mover. Figure on the
  right: incoming right mover.}
  \label{fig:scatt2}
\end{figure}

Let us also consider the scattering of a propagating particle and a bound state. Such processes are depicted on
Fig. \ref{fig:scatt2}. We observe that the single particles move the frozen bound state by 2
lattice units, always in the direction opposite of the particle propagation. In the case of the right mover we observe
that the outgoing orbit is not changed as an effect of the scattering. In the case of the left mover the outgoing orbit
suffers a spatial displacement of $\Delta x=-3$. The outcome of the scattering event does not depend on the length of
the frozen bound state: the length only affects the duration of the scattering. The explanation for this is simply that
as a single particle penetrates the bound state it becomes a hole which can propagate within the bound state with the
same fixed velocity, and the time needed to traverse the bound state depends on its size. The scattering displacements
are summarized in Table \ref{tab:scatt}. 

\begin{table}[b!]
  \small
  \centering
  \begin{tabular}{|c|c||c|c|}
    \hline
    particle coming &   particle coming & $\Delta x_1$ & $\Delta x_2$ \\
    from the left   &  from the right & & \\
    \hline
    \hline
    right mover & left mover & 0  & -1.5\\
    \hline
    right mover & bound state &0  & -2 \\
    \hline
    bound state & left mover & 2&  -3\\
    \hline
  \end{tabular}
  \caption{A summary of the scattering displacements. Here $\Delta x_1$ (or $\Delta x_2$) stands for the spatial displacement of the orbit
  of the particle that was originally incoming from the left (or the right), respectively. The half-integer displacement
  of $-1.5$ in the case of a left mover signals that in this particular case there is a shift of the orbit in both
  spatial and temporal directions, resulting in an average spatial displacement that is half-integer.}
  
  \label{tab:scatt}
\end{table}

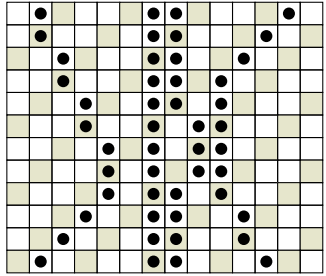
\begin{figure}[b]
  \centering

\begin{tikzpicture}

      
\matrix (m) [matrix of nodes, row sep=-\pgflinewidth, column sep=-\pgflinewidth, 
    nodes={mycell}, nodes in empty cells]
    {
& |[dot]| & |[rec]|  &&&|[rec]|  & |[dot]| &  |[dot]| & |[rec]|  &&&|[rec]|  & |[dot]| & \\ 
& |[dot2]| & &&|[rec]|  && |[dot]| &  |[dot2]| & &&|[rec]|  & |[dot]| & &|[rec]|\\ 
|[rec]|  && |[dot]| & |[rec]|  &&& |[dot2]| &  |[dot]| & &|[rec]|  & |[dot]| & &|[rec]|  &\\ 
&& |[dot2]| & &&|[rec]|  & |[dot]| &  |[dot]| & |[rec]|  & |[dot]| & &|[rec]|  &&\\ 
&|[rec]|  && |[dot]| & |[rec]|  && |[dot]| &  |[dot2]| & & |[dot]| & |[rec]|  &&&|[rec]|\\ 
|[rec]|  &&& |[dot2]| & && |[dot2]| & & |[dot]| &  |[dot2]| & &&|[rec]|  &\\ 
&&|[rec]|  && |[dot]| & |[rec]|  & |[dot]| & & |[dot2]| &  |[dot]| & &|[rec]|  &&\\ 
&|[rec]|  &&& |[dot2]| & & |[dot]| & |[rec]|  & |[dot]| &  |[dot]| & |[rec]|  &&&|[rec]|\\ 
|[rec]|  &&&|[rec]|  & |[dot]| & & |[dot2]| &  |[dot]| & & |[dot2]| & &&|[rec]|  &\\ 
&&|[rec]|  & |[dot]| & &|[rec]|  & |[dot]| &  |[dot]| & |[rec]|  && |[dot]| & |[rec]|  &&\\ 
&|[rec]|  & |[dot]| & &|[rec]|  && |[dot]| &  |[dot2]| & && |[dot2]| & &&|[rec]|\\ 
|[rec]|  & |[dot]| & &|[rec]|  &&& |[dot2]| &  |[dot]| & &|[rec]|  && |[dot]| & |[rec]|  &\\ 
    };
  \end{tikzpicture}
  
  \caption{An example for a three body scattering, with an incoming right mover from the left, an incoming left mover
    from the right, and a bound state of length 2. The displacements are found to be additive: The orbit of the right
    mover is not modified, the position of the bound state is not changed either (displacements due to the left mover
    and right mover add up to zero), but the left mover suffers a total displacement of -4.5 (once again the half integer
    replacement is understood as a combined effect of displacement in the horizontal and vertical directions).}
  \label{fig:3}
\end{figure}

Let us now turn to multi-particle scattering. A non-trivial multi-particle event can happen only if the incoming
particles all have different speed. There are 3 different fixed velocities in the model, thus the only non-trivial case
is the 3-body scattering involving a right mover, a bound state, and a left mover. For such processes we find that the
scattering displacements are additive, which is a classical version of the ``factorized scattering'' from quantum
integrability. A concrete example for the factorization of the 3-body scattering is shown in Fig \ref{fig:3}.

At present we do not have a transparent proof of the factorization, except looking at all the possible scenarios where a
non-trivial 3-body effect could happen. It is clear that any violation of the additivity of the displacements can only
happen if 3 particles with different velocities occupy positions that are at ``interaction distance'' with each other. The
possibilities for such configurations are limited, and in all cases we found that additivity holds. This property is then
extended to all multi-particle scattering events simply because there are only 3 velocities in the model, thus
non-trivial 4 body processes are forbidden by kinematics.

\section{The bond picture: a 3 site model}

\label{sec:3}

An alternative description of the model is achieved by a site-bond transformation that already appeared in the case of
the folded XXZ model in 
\cite{sajat-folded} (see also \cite{folded1,folded2}). The main observation behind the transformation is that the update rules are spin-flip invariant and
they don't change the state of the control bits. Therefore a simplified rule can be applied by focusing on the bonds
between the sites. The idea is to build a new model of length $L$ where the variables are put on the bonds (links)
between the original sites. For each configuration we write down a $\circ$ to a given bond if the two cells of the bond
have the same state, and a $\bullet$ if they have a different state. Periodic boundary conditions on the original
lattice require to have an even number of $\bullet$ states in the bond model. However, it is possible to relax this
condition, and we can build an independent bond model where such a condition is not applied.

It can then be seen that each unitary update step $U^{(4)}(j)$ translates into the following three-site unitary $U^{(3)}(j)$ in the
bond model:
\begin{equation}
  \label{bondu}
  U^{(3)}(j)=P^\circ_{j+1}+P^\bullet_{j+1} \Pe_{j,j+2}.
\end{equation}
Here the middle bit acts as a control bit, which triggers an exchange of the two outer bits. 
The only transition matrix elements correspond to the moves
\begin{center}
  \begin{tikzpicture}
 
\matrix (m) [matrix of nodes, row sep=-\pgflinewidth, column sep=-\pgflinewidth, 
    nodes={mycell}, nodes in empty cells]
    {
&  |[dot]| & |[dot]| \\
};
\node at (1,0) {$\leftrightarrow$};
\begin{scope}[xshift=2cm]
\matrix (m) [matrix of nodes, row sep=-\pgflinewidth, column sep=-\pgflinewidth, 
    nodes={mycell}, nodes in empty cells]
   {
 |[dot]| & |[dot]| &\\
};
\end{scope}

  \end{tikzpicture}
\end{center}
All other configurations are left invariant.

\begin{figure}[b!]
  \centering
  \scalebox{-1}[1]{\includegraphics[scale=0.7]{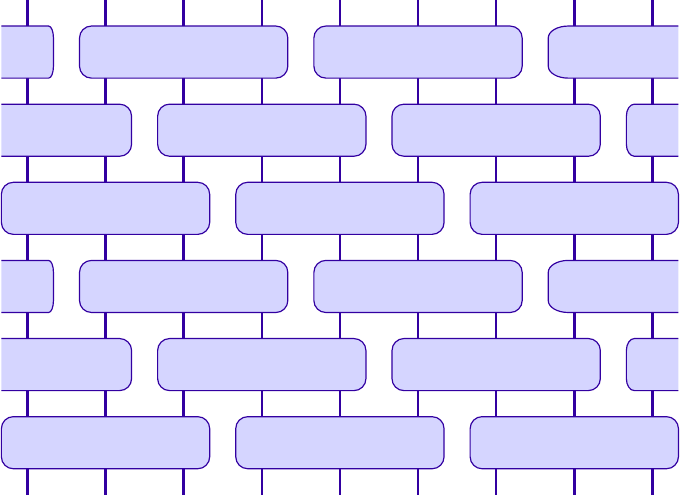}}
  \caption{The system of  gates that are used in the 3-site update rules of the bond model. Each brick has three
  incoming and three outgoing legs, and the action is given by the unitary $U^{(3)}$ defined in the main text. At each
  step the rows
  are shifted to the right by one site. The same brickwork structure is used also in the quantum circuit
  model of Section \ref{sec:gates}.}
  \label{fig:brick1}
\end{figure}

In the bond model the time evolution is constructed as
\begin{equation}
  \label{tV1}
  \tilde{\mathcal{V}}=\tilde{{V}}_3 \tilde{{V}}_2 \tilde{{V}}_1,
\end{equation}
where
\begin{equation}
  \label{tVm}
\tilde  V_m=\prod_{j=1}^{L/3} U^{(3)}(3j+m).
\end{equation}
This time evolution corresponds to a brickwork system of quantum gates depicted in Fig. \ref{fig:brick1}. Notice that
now there is no overlap between the gates, as opposed to the four site unitaries of the original representation, whose
support overlapped at the control bits.

The bond model describes the propagation of particles of finite width 2. Once again we find that there are right movers
with speed 1 and left movers with an average speed of -1/2. The bound states of the original model are now represented
by two single particles that are placed at distance $\ell>1$ from each other. These single particles are interpreted as
``domain walls'', as in the case of the folded XXZ model  \cite{sajat-folded}. The domain walls do not propagate on
their own, but 
they are moved by $\pm 2$ sites as an effect of scattering with a dynamical particle. We do not show separate pictures
for the time evolution in the bond picture, because they can be obtained easily from the graphs of the original
representation by applying the site-bond transformation.

In Section \ref{sec:Lax} we show that the bond model can be embedded into the canonical algebraic framework of
integrability. In order to do this we first introduce a quantum mechanical generalization of the model.

\section{Quantum gate model}

\label{sec:gates}

We consider an extension (or deformation) of the cellular automaton in the bond picture, such that the model becomes
fully quantum 
mechanical. Instead of the deterministic update rule above we introduce a one parameter family of 3-site unitaries given
by 
\begin{equation}
  \label{U3ext}
  U^{(3)}_{j,j+1,j+2}(u)=P^\circ_{j+1}+P^\bullet_{j+1} U^{(2)}_{j,j+2}(u),
\end{equation}
where now the two-site unitary $U^{(2)}_{j,j+2}(u)$ is defined by its explicit matrix form
\begin{equation}
  \label{U2}
  U^{(2)}_{j,j+2}(u)=
  \begin{pmatrix}
    1 & 0 & 0 & 0 \\
      0 & \sech(u) & i\tanh(u) & 0\\
   0 & i\tanh(u)  &  \sech(u)  & 0\\
    0 & 0 & 0 & 1\\
  \end{pmatrix}.
\end{equation}
Here $\sech(u)=1/\cosh(u)$ and the basis states for the tensor product of the spaces $j$ and $j+2$ are ordered in the
standard way of 
$\ket{\circ\circ}, \ket{\circ\bullet}, \ket{\bullet\circ}, \ket{\bullet\bullet}$. The parameter $u$ is assumed to be a
real number. This two-site unitary originates in the $R$-matrix of the XXZ spin chain at the free
fermion point \cite{Korepin-Book}.
Note that the site in the middle still acts as a control bit and its state is not changed during the update step.

Direct computation shows that
\begin{equation}
   U^{(2)}_{j,j+2}(u) U^{(2)}_{j,j+2}(-u)=1,\qquad \left( U^{(2)}_{j,j+2}(u)\right)^\dagger= U^{(2)}_{j,j+2}(-u).
\end{equation}
This confirms the unitarity of $U^{(2)}_{j,j+2}$. The unitarity of the three site gate then follows simply from
eq. \eqref{U3ext}.

The three-site gate $U^{(3)}_{j,j+1,j+2}$ can be used to define a one-parameter family of quantum gate models. The
time evolution is  given by the extension of the formulas \eqref{tV1}-\eqref{tVm} to include 
the parameter dependence:
\begin{equation}
  \label{tVug}
  \begin{split}
     \tilde{\mathcal{V}}(u)&=\tilde{{V}}_3(u) \tilde{{V}}_2(u) \tilde{{V}}_1(u),\\
\tilde  V_m(u)&=\prod_{j=1}^{L/3} U^{(3)}_{3j+m,3j+m+1,3j+m+2}(u).
  \end{split}
\end{equation}
We refer again to Fig. \ref{fig:brick1} for a graphical interpretation.

Our main model, the cellular automaton is reproduced in the $u\to\infty$ limit of the quantum gate model.
Indeed, in this limit we find
\begin{equation}
 \lim_{u\to\infty} U^{(2)}_{j,j+2}(u)=
  \begin{pmatrix}
    1 & 0 & 0 & 0 \\
      0 & 0 & i & 0\\
   0 & i  & 0  & 0\\
    0 & 0 & 0 & 1\\
  \end{pmatrix}.
\end{equation}
This is identical with the permutation matrix up to the phase factors of $i$ appearing in the exchange transition matrix
elements. Thus the three-site unitary \eqref{U3ext} becomes identical with \eqref{bondu} up to these phase factors. It is important, that
this operator is deterministic: if the initial condition 
is a single state from the computational basis, then discrete time evolution will always lead to some other basis state,
multiplied with some phase factor. We can then view this motion as purely classical, because there will not be any
linear combinations and thus the phases become irrelevant. In other words, the update rules for the mean occupation
numbers become identical with the rules of the cellular automaton.

It is also useful to look at the $u\to 0$ limit of the quantum gate model. 
The two-site and three-site unitaries possess the initial conditions
\begin{equation}
  \label{UUinit}
  U^{(2)}_{j,j+2}(0)=1,\qquad U^{(3)}_{j,j+1,j+2}(0)=1.
\end{equation}
Thus at $u=0$ the model becomes completely trivial. The first order Taylor coefficient of the two-site gate gives
\begin{equation}
  U^{(2)}_{j,j+1}(u)=1+iu (\sigma^-_j \sigma^+_{j+1}+\sigma^+_j\sigma^-_{j+1})+\ordo(u^2),
\end{equation}
leading to 
\begin{equation}
  \label{U3exp}
  U^{(3)}_{j,j+1,j+2}(u)=1+iu h_{j,j+1,j+2}+\ordo(u^2),
\end{equation}
with the Hermitian three-site operator
\begin{equation}
h_{j,j+1,j+2}=\sigma^-_j P^\bullet_{j+1}\sigma^+_{j+2}+\sigma^+_j P^\bullet_{j+1}\sigma^-_{j+2}.
\end{equation}
Considering the expansion of the 3-long cycle $\tilde{\mathcal{V}}(u)$ we get
\begin{equation}
  \tilde{\mathcal{V}}(u)=1+iuH+\ordo(u^2),
\end{equation}
where now $H$ is a translationally invariant Hamiltonian
\begin{equation}
 H=\sum_{j=1}^L h_{j,j+1,j+2}.
\end{equation}
This is the Hamiltonian of the ``folded XXZ model''
in the bond model of \cite{sajat-folded} or
in the dual picture of \cite{folded1,folded2}. This model is a special case of the more general Bariev model \cite{bariev-model}.
Thus the quantum gate model can be regarded as a Trotterization of the
folded XXZ and Bariev models. It 
is remarkable, that this family of quantum models includes the cellular automaton as a special case.

It is important that the spectral parameter dependent Floquet operators $\tilde{\mathcal{V}}(u)$ do not form a commuting family:
\begin{equation}
  [ \tilde{\mathcal{V}}(u), \tilde{\mathcal{V}}(v)]\ne 0.
\end{equation}
However, the quantum gate model is still integrable: there is a diagonal-to-diagonal
transfer matrix which belongs to a commuting family; this is shown in the next Section.
Thus the model can be regarded as an
``integrable Trotterization'' of the folded XXZ model, and it can be seen as a 3-site analog of the
construction of \cite{integrable-trotterization}, which is an integrable Trotterization of the XXZ spin chain.

For the sake of completeness we also present the quantum gate generalization of the original 4-site update rule of the
cellular automaton. Instead
of \eqref{U4} we have
\begin{equation}
  U^{(4)}(j|u)=\left(P^\bullet_j P^\bullet_{j+3}+P^\circ_j P^\circ_{j+3}  \right) U^{(2)}_{j+1,j+2}(u)
  +P^\bullet_j P^\circ_{j+3}+P^\circ_j P^\bullet_{j+3} 
\end{equation}
with the same two-site unitary given by \eqref{U2}.
Expanding to first order in $u$ we get
\begin{equation}
    U^{(4)}(j|u)=1+iu h_{j,j+1,j+2,j+3} +\ordo(u^2)
\end{equation}
with the four-site Hermitian operator
\begin{equation}
  h_{j,j+1,j+2,j+3}=\left(P^\bullet_j P^\bullet_{j+3}+P^\circ_j P^\circ_{j+3}  \right)
  \left(\sigma^-_{j+1}\sigma^+_{j+2}+\sigma^+_{j+1}\sigma^-_{j+2}\right).
\end{equation}
Expanding the $u$-dependent generalization of \eqref{V} we get
\begin{equation}
  \mathcal{V}(u)=1+iuH+\ordo(u^2),
\end{equation}
where now
\begin{equation}
  \label{Hbond}
 H=\sum_{j=1}^L h_{j,j+1,j+2,j+3}.
\end{equation}
Apart from a trivial multiplicative normalization this the Hamiltonian of the folded XXZ model treated in
\cite{fracton1,folded1,folded2,sajat-folded}.

\section{Integrability}

\label{sec:Lax}

Now we establish the integrability structure behind the construction, by embedding a certain diagonal-to-diagonal
transfer matrix of the bond model into the canonical framework of integrability. As explained in the previous Section,
our quantum gate model is a Trotterization of the folded XXZ model, which is a special case of the Bariev model. The
algebraic structures behind the Bariev model were worked out in \cite{bariev-lax-1,bariev-lax-2}. Our construction to be
presented below is independent from these works, but the final result is identical to a specialization of the structures
found in \cite{bariev-lax-1,bariev-lax-2}.

\begin{figure}[b!]
  \centering
  \includegraphics[scale=0.7]{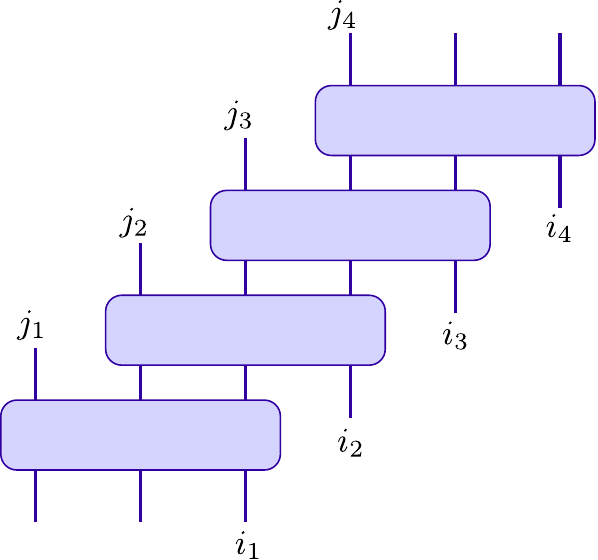}
  \caption{Schematic representation of the diagonal-to-diagonal transfer matrix constructed from three site
    unitaries. The variables $i_1,i_2,\dots$ and  $j_1,j_2,\dots$ represent the incoming and outgoing indices of the
    transfer matrix. There are furthermore 4 additional indices corresponding to the first two incoming variables of the
  first gate, and to the last two outgoing variables of the last gate. Integrability can be established if periodic boundary
conditions are applied to these extra 4 variables, see below in Fig. \ref{fig:tmasik}.} 
  \label{fig:dtod}
\end{figure}

Let us consider the
concatenated action of the $U^{(3)}(u)$ gates with some fixed $u$ along a diagonal direction, following
the one-site shifts in the definition of \eqref{tVug}. Formally we have on a chain of length $L$
\begin{equation}
  \tau(u)=U^{(3)}_{L-2,L-1,L}  \dots   U^{(3)}_{3,4,5}(u)U^{(3)}_{2,3,4}(u)U^{(3)}_{1,2,3}(u).
\end{equation}
We see that the support for each new gate overlaps with that of the preceding gate on exactly two sites. A graphical
representation of such a diagonal-to-diagonal transfer matrix is given in Fig. \ref{fig:dtod}.

The operator $\tau(u)$ is well defined for every finite volume, however, an integrable version is found only in two
special situations: either in
the infinite volume limit or if we consider periodic boundary conditions along a diagonal direction. We focus on  this
second possibility. We distribute the incoming and outgoing variables of the transfer matrix such that it will act
diagonally from bottom right to the top left, and we apply periodic boundary conditions in a special way: we connect the
first two spins at the bottom with the last two spins at the top (see Fig. \ref{fig:dtod}). This is an unphysical
boundary condition because it connects spin variables at different times in the lattice. However, this choice leads to
a relatively simple embedding of the unitary gates to the standard Yang-Baxter framework of integrability. We give
further comments about the utility of the diagonal-to-diagonal transfer matrix at the end of this Section.

Now we present our construction for the transfer matrix. We put forward that the Lax operator and $R$-matrix that we
find turn out to be equal to a special case of the same objects given in \cite{bariev-lax-1,bariev-lax-2} for the Bariev
model. The folded XXZ model corresponds to the special choice $U=\pm 1$ of the Bariev model \cite{bariev-model}, thus it
is very natural that our special solution is included in the general set of results available in the
literature \footnote{The precise relation between our result and those of \cite{bariev-lax-1,bariev-lax-2} was computed
  by Tam\'as Gombor. The details will be presented elsewhere.}. However, we obtained our Lax operator and $R$-matrix using a different route, and for this special model
our formulas take simpler form than those of  \cite{bariev-lax-1,bariev-lax-2}. This is the reason why we chose to
publish our derivations. 

\begin{figure}[b!]
  \centering
  \includegraphics[scale=0.7]{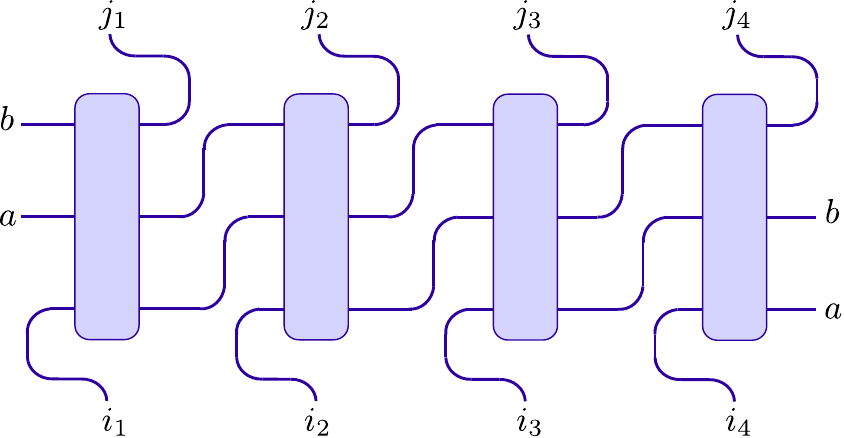}
  \caption{Schematic representation of the transfer matrix \eqref{jot}. Observe that the connectivity of the legs of the
    quantum gates is the same as in Fig. \ref{fig:dtod}, except for the auxiliary spaces $a$ and $b$ where we apply
    periodic boundary conditions. 
    }
  \label{fig:tmasik}
\end{figure}

Let us then consider a tensor product of two auxiliary spaces $a$ and $b$ and a Lax operator $\La_{j,a,b}(u)$ which
acts on the tensor product of a physical space $j$ and the two auxiliary spaces. The action of $\La_{j,a,b}(u)$  is
given by the three-site unitary above and a permutation which takes care of the proper placement of incoming and
outgoing indices for the diagonal-to-diagonal transfer matrix. We get
\begin{equation}
  \label{eq:La}
  \La_{j,a,b}(u)= \Pe_{a,b} \Pe_{j,a} U^{(3)}_{j,a,b}(u).
\end{equation}
Using this Lax operator we define a new transfer matrix acting on $L$ spins as
\begin{equation}
  \label{jot}
  t(u)=\text{Tr}_{a,b}  \left[  \La_{L,a,b}(u)   \dots   \La_{2,a,b}(u)\La_{1,a,b}(u)\right].
\end{equation}
The trace is taken over both auxiliary spaces. A graphical interpretation of this transfer matrix is given
in Fig. \ref{fig:tmasik}.

These transfer matrices belong to a commuting family:
\begin{equation}
  [t(u),t(v)]=0.
  \label{tutv}
\end{equation}
This can be proven using the standard steps of the Quantum Inverse Scattering Approach \cite{Korepin-Book}. Let us take
two pairs of 
auxiliary spaces with labels $a,b$ and $c,d$, and let us label the resulting tensor product  spaces with $A$ and
$B$. Then we can 
express the two 
copies of the transfer matrices as
\begin{equation}
  \begin{split}
    t(u)&=\text{Tr}_{A}  \left[  \La_{L,A}(u)   \dots   \La_{2,A}(u)\La_{1,A}(u)\right]\\
        t(v)&=\text{Tr}_{B}  \left[  \La_{L,B}(v)   \dots   \La_{2,B}(v)\La_{1,B}(v)\right],\\
  \end{split}
 \end{equation}
 where now $\La_{j,A}(u)$ and $\La_{j,B}(v)$ stand for the Lax operator acting on the tensor product of a physical
 space $j$ and the 4 dimensional auxiliary spaces $A$ and $B$. The commutativity
 of the transfer matrices can be established if the Lax operators satisfy the so-called RLL relations \cite{Korepin-Book}
\begin{equation}
  \label{RLL}
  \begin{split}
  R_{B,A}(v,u) & \mathcal{L}_{B,j}(v)  \mathcal{L}_{A,j}(u)= \mathcal{L}_{A,j}(u) \mathcal{L}_{B,j}(v)    R_{B,A}(v,u).
\end{split}
\end{equation}
Here $R_{B,A}(v,u)$ is the so-called $R$-matrix acting on the two  auxiliary spaces $A$ and $B$. The
spaces $A$ and $B$ are 4 dimensional, therefore $R$ is a matrix of size $16\times 16$.

Consistency of the RLL relations requires that the  $R$-matrix satisfies the Yang-Baxter relations, which are operator
equations acting 
on a tensor product of 3 auxiliary spaces:
\begin{equation}
  \label{YB}
  \begin{split}
     R_{12}(u_{1},u_2)&R_{13}(u_1,u_3)R_{23}(u_2,u_3)=\\
&  =R_{23}(u_2,u_3) R_{13}(u_1,u_3) R_{12}(u_{1},u_2).
  \end{split}
\end{equation}

In our case we find that the Lax operators \eqref{eq:La} satisfy the RLL relations with the following $R$-matrix
\footnote{This form of the $R$-matrix was computed by Arthur Hutsalyuk, intended for a separate article to be published in
  collaboration with the present author and others. We are thankful to Arthur Hutsalyuk for letting us use this result
  before the appearance of the upcoming publication.}:
\begin{equation}
\label{R-hrdXXZ}
R_{B,A}(v,u)=
\begin{pmatrix}
E_{11}+E_{44}\rho_1& E_{21}&E_{31}+E_{42}\rho_2&E_{41}\rho_5\\
E_{12}+E_{34}\rho_2& E_{22}+E_{44}\rho_6 &E_{32}\rho_3&E_{31}\rho_4+E_{42}\rho_5\\
E_{13}& E_{23}& E_{33}+E_{44}\rho_6& E_{43}\rho_5\\
E_{14}\rho_5& E_{24}\rho_5& E_{12}\rho_4+E_{34}\rho_5& E_{11}\rho_7+(E_{22}+E_{33})\rho_6+E_{44}
\end{pmatrix}.
\end{equation}
In the above formula the $R$-matrix is written in a block form, such that it is a $4\times 4$ matrix with respect to the
first auxiliary space 
and the matrix elements written there are operators acting on the second auxiliary space. 
The operators $E_{\alpha\beta}$ are the elementary matrices with a single matrix element equal to 1 in the row $\alpha$
and column $\beta$. The rapidity dependent functions in the matrix elements above are 
\begin{equation}
\begin{split}
&\rho_1=\tanh(v-u)\tanh(u),\\
&\rho_2=i\frac{\tanh(v-u)}{\cosh(u)},\\
&\rho_3=\frac{1}{\cosh(v-u)}\left(\frac{\cosh(u)}{\cosh(v)}+\sinh(v-u)\tanh(u)\right),\\
& \rho_4=i\frac{\tanh(v-u)}{\cosh(v)},\\
&\rho_5=\frac{1}{\cosh(v-u)},\\
&\rho_6=i\tanh(v-u),\\
&\rho_7=-\tanh(v-u)\tanh(v).
\end{split}
\end{equation}
This $R$-matrix is of non-difference form and it is not symmetric with respect to the two
spaces. However, it still satisfies the inversion property
\begin{equation}
  R_{B,A}(u,v)R_{A,B}(v,u)=1 \quad \text{ with } \quad R_{A,B}(v,u)= \Pe_{A,B}R_{B,A}(v,u)\Pe_{A,B}.
\end{equation}
It also satisfies the so-called regularity condition
\begin{equation}
  R_{B,A}(u,u)=\Pe_{A,B}.
\end{equation}
The RLL relations and the Yang-Baxter equation can be checked by direct substitution. Afterwards the standard steps of the
QISM \cite{Korepin-Book} can be applied to prove the commutativity \eqref{tutv}.

It is useful to consider the behaviour of $t(u)$ around the initial point $u=0$. The initial condition for the Lax
operator follows from \eqref{eq:La} and \eqref{UUinit}. We get
\begin{equation}
   \La_{j,a,b}(u)= \Pe_{a,b} \Pe_{j,a}.
\end{equation}
This is an unusual initial condition, and to our best knowledge it is new. For the transfer matrix it translates into
\begin{equation}
  t(0)=\UU^2,
\end{equation}
where $\UU$ is the cyclic shift operator on the finite chain of length $L$. Once again, this initial condition appears to
be new: in the standard cases one has simply just $t(0)=\UU$. The fact that the transfer matrix describes translation by
 two sites is very closely tied to the multi-site interactions in the model.

Let us now consider the Taylor expansion of this transfer matrix. Using the expansion \eqref{U3exp} we obtain the first
order terms as
\begin{equation}
  t(u)=\UU^2\left(1+iuH+\ordo(u^2)\right),
\end{equation}
where $H$ is again given by \eqref{Hbond}. Thus the diagonal-to-diagonal transfer matrix also accommodates the ``folded
XXZ model'' (in the bond picture). Due to commutativity of the transfer matrices we can also define further conserved
charges by taking the logarithmic derivatives
\begin{equation}
\left.  \left(\frac{\partial}{\partial u}\right)^k\log(t(u))\right|_{u=0}.
\end{equation}
Simple arguments show that all of these charges are extensive operators with a short range density, completely analogous
to the standard cases. However, the
range of these charge densities grows faster than in nearest neighbour interacting chains. The first
derivative gives the Hamiltonian which is a 3-site operator, and the next charge is actually a 5-site operator; the
range is then increased by 2 with every further derivative.

With this we have obtained an infinite set of local charges for our model. These charges also commute with the Hamiltonian of 
the ``folded XXZ model''.
The Bethe Ansatz solution for the eigenstates is already available, it was already given in
\cite{folded1,folded2,sajat-folded}, which diagonalized the Hamiltonian \eqref{Hbond}. Alternatively, the solution could
be found by a special limit of the computations given in \cite{bariev-model,bariev-lax-1,bariev-lax-2}, which could also
yield the finite volume transfer matrix eigenvalues.

It is interesting that both $t(u)$ and the 3-cycle operator $\mathcal{V}(u)$ involve the Hamiltonian
\eqref{Hbond} in the first order expansion in $u$  around $u=0$. However, the two operators become different at higher orders in
$u$, and only $t(u)$ forms a commuting family. 

Let us now return to the original Cauchy problem of the cellular automaton.
It is clear that the transfer matrix $t(u)$ can not solve the initial value problem, because it has periodic
boundary conditions connecting cells at different time steps. One of the unwanted consequences is that the
diagonal-to-diagonal transfer matrix is completely insensitive to the right movers. Those particles actually propagate
in parallel with the gates of $t(u)$, and thus the right movers appear as a particle with ``infinite velocity'' if viewed from the
``coordinate frame'' of $t(u)$. An analogy is found with relativistic field theories: the right mover is
propagating with the ``speed of light'', and choosing $t(u)$ as the generator of the dynamics corresponds to choosing light
cone coordinates. This is useful only if initial values and boundary values are specified along the light cones, which
is a somewhat different setup.

Nevertheless we believe that the Lax operator construction is a clear sign of the algebraic integrability of our
model. Perhaps a proper generalization of the Lax operators would actually lead to integrable
row-to-row transfer matrices. We leave this as an open problem for future research.

\section{Discussion}

\label{sec:conclude}

We presented a new cellular automaton with two different formulations. In the ``original picture'' the update rule uses
two control bits and two action bits, whereas in the ``bond picture'' only one control bit is used  with two action
bits. It is important to summarize the connections with existing models in the literature.

First of all, our model can be regarded as a generalization of the Rule54 model, but there are also key differences.
Our model has three particle types with three
different velocities (left movers, right movers and bound states), and this is richer than simply the left- and right-
movers of the Rule54 model. Perhaps our model is the simplest one where factorization of the multi particle
scattering can be observed. In the Rule54 model there are no 3 body collisions, but our model has room for such
configurations, see for example the scattering event depicted on Fig. \ref{fig:3}.
 
Our model is also similar to various quantum circuit models that appeared in the literature. As explained in Section \ref{sec:gates}
it can be considered an integrable Trotterization of the ``folded XXZ model'', and in this respect it is a 3-site (or
4-site) generalization of the work \cite{integrable-trotterization}. Our model is also similar to the dual unitary
gate models of \cite{dual-unitary-0,prosen-dual1,dual-unitary-2} and the dual unitary round-a-face models of
\cite{prosen-round-a-face}. These models apply one control bit (dual unitary gate models) or two control bits (dual unitary
round-a-face models) for the action of a one site
unitary. In our model we have one or two control bits (depending on the formulation) for the action of a two site
unitary. Thus we obtained the first member of an even wider class of quantum gate models, potentially including
even more integrable models.

One of our key results is the construction of the Lax operator \eqref{eq:La}, which has two new
properties as opposed to the standard cases in the literature: it uses two auxiliary spaces (which are coupled), and its
initial value at zero rapidity is the cyclic permutation operator acting on the  physical space  and the two auxiliary spaces. As an
effect, the initial value for the resulting transfer matrix is the two-site cyclic shift operator for the underlying
spin chain. This leads to a 3-site Hamiltonian, and an infinite family of conserved charges, such that there is no
2-site operator commuting with them. Nevertheless the charges are translationally invariant, with respect to single
site shifts. Similar constructions already appeared in the literature. The most important example is the algebraic
treatment of the Bariev model
\cite{bariev-lax-1,bariev-lax-2}, which actually includes our solution in a special limit.
Other examples for similar constructions are zig-zag spin ladders or
coupled XXZ spin chains (see for example \cite{zigzag1a,zigzag2,zigzag3}). Nevertheless the formalism presented in this
paper seems to be new, and for this particular model our formulas are more transparent than those of
\cite{bariev-lax-1,bariev-lax-2}.

Let us now discuss some of the open questions. First of all, it would be important to find out whether the row-to-row
transfer matrix $\mathcal{V}(u)$ possesses any integrability properties. Even though $[\mathcal{V}(u),\mathcal{V}(v)]\ne
0$ it is still possible that there is a multi-parametric extension of these transfer matrices, such
that a restricted commutativity would still hold within the larger family. This would then extend the commutativity
properties of the transfer matrices in \cite{integrable-trotterization}. However, at present it is not clear whether
such a structure should be expected, and it would be desirable to study the level statistics $\mathcal{V}(u)$ to get a
working hypothesis about its integrability. We remind that the diagonal-to-diagonal transfer matrix $t(u)$ can not
handle the Cauchy problem of the cellular automaton (see the discussion at the end of the previous Section), and this
motivates further study of the row-to-row transfer matrix $\mathcal{V}(u)$.

It would be interesting to determine the large scale transport properties of the present models. We expect to find
ballistic transport with diffusive corrections, just like in the Rule54 model
\cite{rule54-transport,sarang-vasseur-huse}.
As argued above, our cellular automaton is perhaps the simplest model with factorized
multi-particle scattering, therefore it could be used as a further testing ground for
the predictions of GHD.

Finally, perhaps the most interesting question is, what type of new integrable models we can still obtain with the present
algebraic construction. In this work we used one or two control bits and the $R$-matrix of the XX chain for the two-site
unitary $U^{(2)}(u)$. An obvious idea is to use the known $R$-matrix of the XXZ chain for
the same purpose. This will not give us a new cellular automaton, but it will result in new integrable spin chains (or
quantum gate models), which can
be considered as ``hard rod deformations'' of the XXZ spin chains. This will be presented in a separate publication in
collaboration with other researchers.

\vspace{1cm}
{\bf Acknowledgments} 

\bigskip

We are grateful to Tam\'as Gombor, Arthur Hutsalyuk, Yunfeng Jiang, Toma\v{z} Prosen, Eric Vernier and Lenart Zadnik for
useful 
discussions. In 
particular we thank Arthur Hutsalyuk for providing us the $R$-matrix in eq. \ref{R-hrdXXZ}, and Tam\'as Gombor for
checking some of the formulas and for his help with some of the Figures in this document. We are grateful to Lenart
Zadnik for referring us to the papers \cite{bariev-lax-1,bariev-lax-2} which treated the algebraic structures behind the
Bariev model.

\addcontentsline{toc}{section}{References}

\begin{thebibliography}{10}

\bibitem{cellaut-review}
M.~Mitchell, {\em Computation in Cellular Automata: A Selected Review},
  \href{http://dx.doi.org/https://doi.org/10.1002/3527602968.ch4}{ch.~4,
  pp.~95--140}.
\newblock John Wiley {\&} Sons, Ltd, 1998.

\bibitem{wolfram-book}
S.~Wolfram, {\em A New Kind of Science}.
\newblock Wolfram Media, 2002.

\bibitem{rendell-game-of-life-book}
P.~Rendell, \href{http://dx.doi.org/10.1007/978-3-319-19842-2}{{\em Turing
  Machine Universality of the Game of Life}}.
\newblock Springer International Publishing, 2016.

\bibitem{wolfram-cellaut1}
S.~Wolfram, ``Statistical mechanics of cellular automata,''
  \href{http://dx.doi.org/10.1103/RevModPhys.55.601}{{\em Rev. Mod. Phys.} {\bf
  55} (1983)  601--644}.

\bibitem{cellaut-class}
S.~Takesue, ``Reversible Cellular Automata and Statistical Mechanics,''
  \href{http://dx.doi.org/10.1103/PhysRevLett.59.2499}{{\em Phys. Rev. Lett.}
  {\bf 59} (1987)  2499--2502}.

\bibitem{rule54}
A.~Bobenko, M.~Bordemann, C.~Gunn, and U.~Pinkall, ``{On two integrable
  cellular automata},'' \href{http://dx.doi.org/10.1007/BF02097234}{{\em Comm.
  Math. Phys.} {\bf 158} (1993) no.~1, 127 -- 134}.

\bibitem{sutherland-book}
B.~Sutherland, {\em Beautiful Models}.
\newblock World Scientific Publishing Company, 2004.

\bibitem{Baxter-Book}
R.~J. Baxter, {\em Exactly solved models in statistical mechanics}.
\newblock London: Academic Press Inc, 1982.

\bibitem{mussardo-review}
G.~Mussardo, ``{Off-critical statistical models: Factorized scattering theories
  and bootstrap program},''
\href{http://dx.doi.org/10.1016/0370-1573(92)90047-4}{{\em Phys. Rept.} {\bf
  218} (1992)  215--379}.

\bibitem{doyon-ghd}
O.~A. {Castro-Alvaredo}, B.~{Doyon}, and T.~{Yoshimura}, ``{Emergent
  Hydrodynamics in Integrable Quantum Systems Out of Equilibrium},''
  \href{http://dx.doi.org/10.1103/PhysRevX.6.041065}{{\em Phys. Rev. X} {\bf 6}
  (2016) no.~4, 041065}, \href{http://arxiv.org/abs/1605.07331}{{\tt
  arXiv:1605.07331 [cond-mat.stat-mech]}}.

\bibitem{jacopo-ghd}
B.~{Bertini}, M.~{Collura}, J.~{De Nardis}, and M.~{Fagotti}, ``{Transport in
  Out-of-Equilibrium XXZ Chains: Exact Profiles of Charges and Currents},''
  \href{http://dx.doi.org/10.1103/PhysRevLett.117.207201}{{\em Phys. Rev.
  Lett.} {\bf 117} (2016) no.~20, 207201},
  \href{http://arxiv.org/abs/1605.09790}{{\tt arXiv:1605.09790
  [cond-mat.stat-mech]}}.

\bibitem{ghd-misc}
A collection of review articles in the topic of GHD will appear in the near
  future in a Special Issue of J. Stat. Mech.

\bibitem{doyon-hydro-proj}
B.~{Doyon}, ``{Hydrodynamic projections and the emergence of linearised Euler
  equations in one-dimensional isolated systems},'' {\em arXiv e-prints} (2020)
   , \href{http://arxiv.org/abs/2011.00611}{{\tt arXiv:2011.00611 [math-ph]}}.

\bibitem{granet-essler-ghd}
E.~{Granet} and F.~H.~L. {Essler}, ``{Systematic strong coupling expansion for
  out-of-equilibrium dynamics in the Lieb-Liniger model},'' {\em arXiv
  e-prints} (2021)  , \href{http://arxiv.org/abs/2102.09987}{{\tt
  arXiv:2102.09987 [cond-mat.stat-mech]}}.

\bibitem{sajat-currents-review}
M.~{Borsi}, B.~{Pozsgay}, and L.~{Pristy{\'a}k}, ``{Current operators in
  integrable models: A review},'' {\em arXiv e-prints} (2021)  ,
  \href{http://arxiv.org/abs/2103.12160}{{\tt arXiv:2103.12160
  [cond-mat.stat-mech]}}.

\bibitem{katja-bruno-rule54-ghd}
K.~Klobas and B.~Bertini, ``{Exact relaxation to Gibbs and non-equilibrium
  steady states in the quantum cellular automaton Rule 54},'' {\em arXiv
  e-prints} (2021)  , \href{http://arxiv.org/abs/2104.04511}{{\tt
  arXiv:2104.04511 [cond-mat.stat-mech]}}.

\bibitem{rule54-review}
B.~{Bu{\v{c}}a}, K.~{Klobas}, and T.~{Prosen}, ``{Rule 54: Exactly solvable
  model of nonequilibrium statistical mechanics},'' {\em arXiv e-prints} (2021)
   , \href{http://arxiv.org/abs/2103.16543}{{\tt arXiv:2103.16543
  [cond-mat.stat-mech]}}.

\bibitem{katja-bruno-lorenzo-rule54}
K.~Klobas, B.~Bertini, and L.~Piroli, ``Exact Thermalization Dynamics in the
  ``Rule 54'' Quantum Cellular Automaton,''
  \href{http://dx.doi.org/10.1103/PhysRevLett.126.160602}{{\em Phys. Rev.
  Lett.} {\bf 126} (2021)  160602}, \href{http://arxiv.org/abs/2012.12256}{{\tt
  arXiv:2012.12256 [cond-mat.stat-mech]}}.

\bibitem{rule54-entangl}
K.~{Klobas} and B.~{Bertini}, ``{Entanglement dynamics in Rule 54: exact
  results and quasiparticle picture},'' {\em arXiv e-prints} (2021)  ,
  \href{http://arxiv.org/abs/2104.04513}{{\tt arXiv:2104.04513
  [cond-mat.stat-mech]}}.

\bibitem{vasseur-rule54}
A.~J. {Friedman}, S.~{Gopalakrishnan}, and R.~{Vasseur}, ``Integrable Many-Body
  Quantum Floquet-Thouless Pumps,''
  \href{http://dx.doi.org/10.1103/PhysRevLett.123.170603}{{\em Phys. Rev.
  Lett.} {\bf 123} (2019)  170603}, \href{http://arxiv.org/abs/1905.03265}{{\tt
  arXiv:1905.03265 [cond-mat.stat-mech]}}.

\bibitem{dual-unitary-0}
P.~{Kos}, M.~{Ljubotina}, and T.~{Prosen}, ``{Many-Body Quantum Chaos: Analytic
  Connection to Random Matrix Theory},''
  \href{http://dx.doi.org/10.1103/PhysRevX.8.021062}{{\em Phys. Rev. X} {\bf 8}
  (2018) no.~2, 021062}, \href{http://arxiv.org/abs/1712.02665}{{\tt
  arXiv:1712.02665 [nlin.CD]}}.

\bibitem{prosen-dual1}
B.~Bertini, P.~Kos, and T.~Prosen, ``Exact Correlation Functions for
  Dual-Unitary Lattice Models in 1+1 Dimensions,''
  \href{http://dx.doi.org/10.1103/physrevlett.123.210601}{{\em Phys. Rev.
  Lett.} {\bf 123} (2019) no.~21, },
  \href{http://arxiv.org/abs/1904.02140}{{\tt arXiv:1904.02140
  [cond-mat.stat-mech]}}.

\bibitem{dual-unitary-2}
L.~{Piroli}, B.~{Bertini}, J.~I. {Cirac}, and T.~{Prosen}, ``{Exact dynamics in
  dual-unitary quantum circuits},''
  \href{http://dx.doi.org/10.1103/PhysRevB.101.094304}{{\em Phys. Rev. B} {\bf
  101} (2020) no.~9, 094304}, \href{http://arxiv.org/abs/1911.11175}{{\tt
  arXiv:1911.11175 [cond-mat.stat-mech]}}.

\bibitem{integrable-trotterization}
M.~{Vanicat}, L.~{Zadnik}, and T.~{Prosen}, ``{Integrable Trotterization: Local
  Conservation Laws and Boundary Driving},''
  \href{http://dx.doi.org/10.1103/PhysRevLett.121.030606}{{\em Phys. Rev.
  Lett.} {\bf 121} (2018) no.~3, 030606},
  \href{http://arxiv.org/abs/1712.00431}{{\tt arXiv:1712.00431
  [cond-mat.stat-mech]}}.

\bibitem{fracton1}
Z.-C. {Yang}, F.~{Liu}, A.~V. {Gorshkov}, and T.~{Iadecola}, ``{Hilbert-Space
  Fragmentation from Strict Confinement},''
  \href{http://dx.doi.org/10.1103/PhysRevLett.124.207602}{{\em Phys. Rev.
  Lett.} {\bf 124} (2020) no.~20, 207602},
  \href{http://arxiv.org/abs/1912.04300}{{\tt arXiv:1912.04300
  [cond-mat.str-el]}}.

\bibitem{folded1}
L.~{Zadnik} and M.~{Fagotti}, ``{The Folded Spin-1/2 XXZ Model: I.
  Diagonalisation, Jamming, and Ground State Properties},''
  \href{http://dx.doi.org/10.21468/SciPostPhysCore.4.2.010}{{\em SciPost Phys.
  Core} {\bf 4} (2021)  10}, \href{http://arxiv.org/abs/2009.04995}{{\tt
  arXiv:2009.04995 [cond-mat.stat-mech]}}.

\bibitem{folded2}
L.~{Zadnik}, K.~{Bidzhiev}, and M.~{Fagotti}, ``{The Folded Spin-1/2 XXZ Model:
  II. Thermodynamics and Hydrodynamics with a Minimal Set of Charges},''
  \href{http://dx.doi.org/10.21468/SciPostPhys.10.5.099}{{\em SciPost Phys.}
  {\bf 10} (2021)  99}, \href{http://arxiv.org/abs/2011.01159}{{\tt
  arXiv:2011.01159 [cond-mat.stat-mech]}}.

\bibitem{sajat-folded}
B.~{Pozsgay}, T.~{Gombor}, A.~{Hutsalyuk}, Y.~{Jiang}, L.~{Pristy{\'a}k}, and
  E.~{Vernier}, ``{An integrable spin chain with Hilbert space fragmentation
  and solvable real time dynamics},'' {\em arXiv e-prints} (2021)  ,
  \href{http://arxiv.org/abs/2105.02252}{{\tt arXiv:2105.02252
  [cond-mat.stat-mech]}}.

\bibitem{bariev-model}
R.~Z. Bariev, ``Integrable spin chain with two- and three-particle
  interactions,'' \href{http://dx.doi.org/10.1088/0305-4470/24/10/010}{{\em
  Journal of Physics A: Mathematical and General} {\bf 24} (1991) no.~10,
  L549--L553}.

\bibitem{Korepin-Book}
V.~Korepin, N.~Bogoliubov, and A.~Izergin, {\em Quantum inverse scattering
  method and correlation functions}.
\newblock Cambridge University Press, 1993.

\bibitem{bariev-lax-1}
H.-Q. Zhou, ``Quantum integrability for the one-dimensional Bariev chain,''
  \href{http://dx.doi.org/https://doi.org/10.1016/0375-9601(96)00521-X}{{\em
  Phys. Lett. A} {\bf 221} (1996) no.~1, 104--108}.

\bibitem{bariev-lax-2}
M.~Shiroishi and M.~Wadati, ``Integrability of the one-dimensional Bariev
  model,'' \href{http://dx.doi.org/10.1088/0305-4470/30/4/014}{{\em J. Phys. A}
  {\bf 30} (1997) no.~4, 1115--1133}.

\bibitem{prosen-round-a-face}
T.~{Prosen}, ``{Many Body Quantum Chaos and Dual Unitarity Round-a-Face},''
  {\em arXiv e-prints} (2021)  , \href{http://arxiv.org/abs/2105.08022}{{\tt
  2105.08022 [cond-mat.stat-mech]}}.

\bibitem{zigzag1a}
V.~Popkov and A.~Zvyagin, ``“Antichiral” exactly solvable effectively
  two-dimensional quantum spin model,''
  \href{http://dx.doi.org/https://doi.org/10.1016/0375-9601(93)90624-9}{{\em
  Phys. Lett. A} {\bf 175} (1993) no.~5, 295--298}.

\bibitem{zigzag2}
H.~{Frahm} and C.~{R{\"o}denbeck}, ``{Integrable models of coupled Heisenberg
  chains},'' \href{http://dx.doi.org/10.1209/epl/i1996-00302-7}{{\em Europhys.
  Lett.} {\bf 33} (1996) no.~1, 47--52},
  \href{http://arxiv.org/abs/cond-mat/9502090}{{\tt arXiv:cond-mat/9502090
  [cond-mat]}}.

\bibitem{zigzag3}
A.~A. Zvyagin, ``Bethe ansatz solvable multi-chain quantum systems,''
  \href{http://dx.doi.org/10.1088/0305-4470/34/41/201}{{\em J. Phys. A} {\bf
  34} (2001) no.~41, R21--R53}.

\bibitem{rule54-transport}
K.~Klobas, M.~Medenjak, T.~Prosen, and M.~Vanicat, ``Time-dependent matrix
  product ansatz for interacting reversible dynamics,''
  \href{http://dx.doi.org/10.1007/s00220-019-03494-5}{{\em Comm. Math. Phys.}
  {\bf 371} (2019) no.~2, 651–688},
  \href{http://arxiv.org/abs/1807.05000}{{\tt arXiv:1807.05000}}.

\bibitem{sarang-vasseur-huse}
S.~{Gopalakrishnan}, D.~A. {Huse}, V.~{Khemani}, and R.~{Vasseur},
  ``{Hydrodynamics of operator spreading and quasiparticle diffusion in
  interacting integrable systems},''
  \href{http://dx.doi.org/10.1103/PhysRevB.98.220303}{{\em Phys. Rev. B} {\bf
  98} (2018) no.~22, 220303}, \href{http://arxiv.org/abs/1809.02126}{{\tt
  arXiv:1809.02126 [cond-mat.stat-mech]}}.

\end{thebibliography}
\providecommand{\href}[2]{#2}\begingroup\raggedright\endgroup

\end{document}